\documentstyle[12pt]{article}
\input epsf
\begin{document}
\addtolength{\baselineskip}{0.5\baselineskip}
\rightline{CALT-68-1904}
\vskip 2cm
\begin{center}
{\large\bf Fermionic $\theta$ Vacua and Long-Necked Remnants\footnote{
Work supported in part by D.O.E. Grant No. DE-FG03-92-ER40701}}
\vskip 1cm
Piljin Yi\footnote{e-mail: piljin@theory.caltech.edu} \\
{\em 452-48 California Institute of Technology, Pasadena, CA 91125, U.S.A.}
\end{center}
\vskip 3cm
\centerline{ABSTRACT}
\vskip 0.5 cm
\begin{quote}
We study a vacuum polarization effect in the background of a certain dilatonic
extremal black hole, known as the cornucopion. Whenever charged fermions
are of any nonzero mass, the gravitational backreaction to a generic value
of a $CP$ nonconserving vacuum angle $\theta$ is shown to be important
owing to a vacuum energy density which does not vanish deep inside the
the cornucopion. When the vacuum energy density is positive, this effect
creates an extremal horizon at finite physical distance, closing off the
infinite neck. We study the geometry of this horizon in some detail and
find different physical interpretations for small and large fermion mass.
Also, we argue that the conclusion is qualitatively correct
despite the inevitable strong coupling.
\end{quote}
\vskip 2cm
\leftline{Revised, February 1994}
\newpage
\leftline{\bf 1. Motivation}
\vskip 0.5cm

It is well-known that a magnetic monopole carries fractional electric
charge in the presence of a $CP$ nonconserving angle $\theta$ \cite{witten}.
For a spontaneouly broken Yang-Mills theory (with monopoles as solitons),
this  angle parametrizes a continuum of  vacua,
known as $\theta$ vacua, the effect of which is naturally incorporated by
including the Pontryagin density multiplied by $\theta$ in the Lagrangian.
With such a setup, the origin of the fractional charge becomes quite clear.
The Pontryagin density is essentially a product of electric and magnetic
fields, through which a classical magnetic field acts as a source to
the fluctuating electric field. As a result a monopole carries a long
range electric field proportional to its magnetic field, and the
corresponding electric charge must be proportional to $\theta$.
Allowing higher excitations, we arrive at the following Witten's
quantization rule for unit magnetic monopoles.
\begin{equation}
q=N-\frac{\theta}{2\pi},\qquad \hbox{$N$ is any integer.} \label{eq:witten}
\end{equation}
As usual, the quantization rule tells us nothing about how the electric
charge should be realized in such dyons, which must depend on many details
of the theory. One example where we can address this question of dyon core
structure is a monopole coupled to charged fermions
\cite{callan}\cite{rubakov}\cite{dirac}. The lowest partial waves
of such fermions, known to experience no potential barrier, can be used
to study the static dyonic core structure.

As demonstrated by Callan \cite{callan}, the vacuum fluctuation of the
charged fermion field tries to shield the core from the radial electric field
and, as a result, the fractional electric charge is effectively realized as a
vacuum polarization cloud of size $\sim 1/m_{\psi}$ around the monopole,
where $m_{\psi}$ is the mass of the fermion. In particular, for small fermion
mass, the electric charge distribution is concentrated on a thick and large
shell of radius $\sim 1/m_{\psi}$ and the small magnetic core, being shielded
from the extra radial electric field, is found to be essentially intact.

Witten's charge quantization is a topological statement,
and the same fractional charge must appear also for magnetic black holes.
On the other hand, when the magnetic black hole is much smaller than the
fermion lengthscale $\sim 1/m_{\psi}$, we may also expect to find similar
dilute clouds of vacuum polarization shielding the black hole from the extra
radial electric field so that its geometry near the horizon is that of the
pure magnetic black hole.

As far as light fermions are concerned, the leading effect in such
a background is from the long-range magnetic field, and once the ``core''
region is shielded by the resulting dilute cloud, one may argue, there is
little to which  gravity reacts. While the horizon can affect
the dynamics of the fermion fields nearby, this seems to matter only
when the fermion fields are massive enough for the
charged cloud to approach the event horizon.

However, there is a curious species of extremal magnetic black holes, known
as cornucopia, whose ``size'' is both small and infinite simultaneously.
Instead of a compact core, the cornucopion has an infinitely long neck with
the transverse radius proportional to the total magnetic flux, as illustrated
in Figure 1. It takes literally infinite proper time just to reach the event
horizon at the bottom of the neck, let alone to cross it, while the
transverse size of the neck can be arbitrarily small.

\vskip 15mm
\begin{center}
\leavevmode
\epsfysize=2.5in \epsfbox{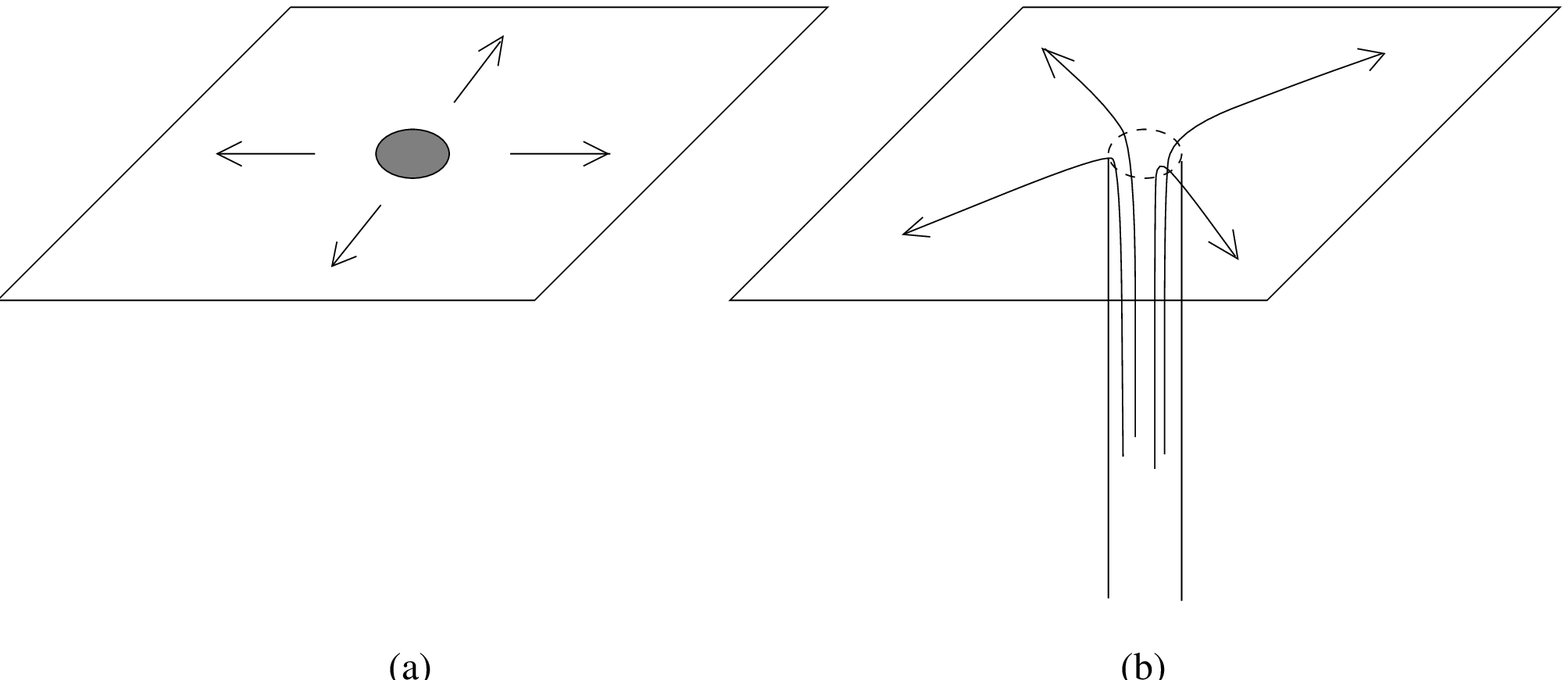}
\end{center}
\vskip 5mm
\begin{quote}
{\bf Figure 1:} {\small Schematic diagrams for (a)
a magnetic monopole in a flat
space-time, and for (b) a cornucopion in an asymptotically flat space-time.
Magnetic flux emanating from the central region is denoted by the arrows.
A cornucopion, which is an extremally charged dilatonic black hole, has an
infinitely long neck of fixed transverse radius threaded by the
magnetic flux.}
\end{quote}
\vskip 10mm

Then we may ask how the vacuum polarization behaves in such an exotic
background. Should one expect to find the narrow and {\it infinite} neck
surrounded by a large and dilute charged cloud of vacuum fluctuation, provided
that the fermion mass is small enough?  After all, far away from the black
hole, there is little indication as to the existence of the infinite neck.
In this paper, we want to address this specific question, by studying
a bosonized effective action for the S-wave charged fermions coupled to
the dilaton gravity.

After the derivation of the effective action for a general spherically
symmetric background in section 2, we shall return to the specific case of
the cornucopion. The surprising result of section 3 is that the
energy cost of the vacuum polarization, which should be balanced against
the gain in the electrostatic energy, is actually divergent in such a
noncompact background.
The inevitable conclusion thereof is that the gravitational backreaction
to this vacuum polarization process is never negligible and, for whatever
$m_{\psi}\neq 0$ is, must have the characteristic long neck terminated,
by creating an extremal horizon at  finite physical distance.

In section 4, we study the effective action of S-wave fermions combined
with the dilaton gravity in four dimension to investigate the self-consistent
geometries with finite ADM masses. We find a useful and practical way of
studying the solutions near the extremal horizon formed by the gravitational
backreaction, and, using this method, we clarify the different
roles the fractional electric charge plays in large and small fermion
mass limit. In particular, we find the expected behaviour in the large fermion
mass limit, where most of the fractional charge must be trapped by
the black hole's gravitational pull. We conclude by discussing the generality
of these results.

\vskip 0.5cm
\leftline{\bf 2. Callan-Rubakov Modes in a Magnetic Black Hole Background}
\vskip 0.5cm

To be definite, let us consider a static dilatonic black hole solution
with magnetic charge \cite{dilaton}, written down in terms of the tortoise
coordinate $z$.
\begin{equation}
 g=\lambda^2(z) dt^2-\lambda^2(z) dz^2-R^2(z) d\Omega^2,\qquad
 e^{-2\phi}=e^{-2\phi(z)}.
\end{equation}
Asymptotically $R \simeq z \rightarrow \infty$, while the event horizon
is at $z=-\infty$ where the geometry is largely determined by the behaviour
of $\lambda$. For a cornucopion which is a purely magnetic black hole,
we can take $\lambda \equiv 1$. For black holes with an extremal
horizon at finite physical distance, such as those with both electric
and magnetic charges inside the event horizon \cite{kallosh},
$\lambda^2 \sim 1/z^2$
as we approach the horizon. Finally $\phi$ is the dilaton field and
$e^{\phi}$ plays the role of the coupling.  Since we want to couple the
matter system to gravity later on, we shall keep $\lambda$, $R$ and
$\phi$ unspecified for a while.
As long as the solution is static and spherically symmetric, the detailed
form of it does not enter the derivation in this section.

\vskip 5mm
Now depending on the origin of the magnetic charge,
we can introduce different kinds of charged fermions. The simplest case
would be a Dirac fermion coupled to the $U(1)$ gauge field.
However, we found it advantageous to work with a spontaneously broken
$SU(2)$ theory so that the fermions are in the fundamental representation
of $SU(2)$ and that the magnetic charge is realized as a topological
quantum number of solitons. The background gauge field outside the horizon is
still Abelian except that the Abelian $U(1)$
generator $Q$ is expressed in terms of the $SU(2)$ generators $T_{a}$.
\[ Q\equiv T_{a} n^{a},\qquad   \hbox{$\vec{n}$ is the unit radial vector
field.} \]
With this choice, the derivation of the effective matter action is quite
similar to that of Callan~\cite{callan}, up to the conventions regarding the
spinor and the modifications due to the nontrivial geometry.
We shall compare our results to those of Callan whenever appropriate.

\vskip 5mm
Consider a $SU(2)$ doublet Dirac fermion $\psi$ with nonzero mass $m_{\psi}$.
As mentioned in the previous section, it is sufficient to focus on
the lowest partial waves of the fermion, called Callan-Rubakov modes, which
do not see any potential barrier of the geometry or of the spherically
symmetric gauge field \cite{callan}\cite{holzhey}. To isolate such modes,
we use the following ansatz for $\psi_{\pm}$, positive and negative chiral
eigenstates of $\psi$ written in terms of Weyl 2-spinors.
\begin{equation}
\psi_{+} =\frac{1}{\sqrt{4\pi\lambda} R} \chi_{+}(t,z),\qquad
\psi_{-} =\frac{1}{\sqrt{4\pi\lambda} R} \gamma^{t} \chi_{-}(t,z).
\end{equation}
The upper (lower) component of the two spinor $\chi_{-}$ ($\chi_{+}$)
has charge $1/2$ while the other has $-1/2$, with respect to the
unbroken $U(1)$ generator $Q$. We chose $\gamma^{t}=\sigma_{x}$ and
$\gamma^{z}=i\sigma_{y}$ as our two-dimensional Dirac matrices.
With this ansatz, the fermion action can be reduced to the following
two-dimensional form.
\begin{eqnarray}
S_{\chi}&=&\int dt dz\,
      (i \overline{\chi}_{+}\gamma^{i} \partial_{i} \chi_{+}
      +i \overline{\chi}_{-}\gamma^{i} \partial_{i} \chi_{-}
      +m_{\psi}\lambda\,(\overline{\chi}_{+}\chi_{-}+
      \overline{\chi}_{-}\chi_{+}))   \nonumber  \\
     &+&\int dtdz \,4\pi \lambda^2 R^2\, (a_{t}J^{t}_{Q}+a_{z}J^{z}_{Q}),
\end{eqnarray}
where $\vec{a}$ is the fluctuating part of the radial $U(1)$ gauge field.
We can set $a_{t}$ equal to zero using the gauge degree of freedom
and then the radial electric field in $(t,z)$ coordinates is simply
$E \equiv\partial_{t}a_{z}$. The relevant currents are,
\begin{equation}
J^{t}_{Q}=\frac{1}{8\pi\lambda^2 R^2}(J^{z}_{+}-J^{z}_{-}), \qquad
J^{z}_{Q}=\frac{1}{8\pi\lambda^2 R^2}(-J^{t}_{+}+J^{t}_{-}),
\end{equation}
where $J_{\pm}$ are the two-dimensional vector currents of $\chi_{\pm}$.
The effective action $S_{\chi}$ above is incomplete since we neglected the
action for the fluctuating electric field $E$ so far. By isolating
it from the full Yang-Mills action and integrating the angular part,
we find
\begin{equation}
S_{E}=\int dt dz\, (\,\frac{\theta}{2\pi}E+
                     \frac{R^2}{2\lambda^2 e^{2\phi}}E^2).
\end{equation}
The $\theta$ term is from the Pontryagin density and can be deduced from
the fact that a unit magnetic monopole carries total magnetic flux $4\pi$.

This effective action $S_{\chi}+S_{E}$ is different from
that of Callan \cite{callan} in two respects.
First, the effective couplings are changed due to the nontrivial
geometry and the nonuniform coupling $e^{\phi}$. In particular,
the fermion mass term acquires a factor of $\lambda$.
Second, the radial coordinate $z$ extends from $\infty$
to all the way to $-\infty$. Because of this, we no longer need to
impose a boundary condition at the origin.
In fact, it is effectively a theory of fermions in flat 1+1
Minkowski space-time, coupled to a $U(1)$ field through an axial
current, but with position-dependent mass and coupling.

\vskip 0.5cm
Since the $U(1)$ field is coupled to a two-dimensional {\em axial}
current\footnote{The two spinors $\chi_{\pm}$ are of opposite charges,
so that the $U(1)$ gauge symmetry is not anomalous.}, we can bosonize
$\chi_{\pm}$ through  the following fundamental  relationships between
currents \cite{callan}\cite{boson}, preserving the $U(1)$ current
automatically,
\begin{equation}
\vec{J}_{\pm}= -\frac{1}{\sqrt{\pi}}\vec{\partial} f_{\pm},\qquad
\hbox{with respect to the flat metric}\quad dt^2-dz^2.
\end{equation}
Furthurmore, to separate the charged and the uncharged sectors, it is
convenient to perform a canonical transformation generated from
\begin{equation}
f\equiv {(f_{+}-f_{-})}/{\sqrt{2}}, \qquad
\eta\equiv \eta(-\infty)+{\int^{z}_{-\infty}(\dot{f}_{+}+\dot{f}_{-})}
/{\sqrt{2}}.
\end{equation}
Once this is done, we can simply eliminate the electric field strength
$E$ through its equation of motion and express the effective action
completely in terms of bosonic  fields $f$ and $\eta$. If we define
$\mu$ to be the geometric mean of the normal ordering scales
$\mu_{f}$ and $\mu_{\eta}$ for each field, the effective action
$S_{\chi}+S_{E}$ is transformed into
\begin{equation}
\int dtdz\:\biggl\{\frac{1}{2}(\partial f)^2
         +\frac{1}{2}(\partial \eta)^2-\frac{e^{2\phi}\lambda^2}{4\pi R^2}
         (f-\frac{\theta}{\sqrt{2\pi}})^2 +cm_{\psi}\mu\lambda\,
         \cos\sqrt{2\pi}f \cos\sqrt{2\pi}\eta \biggr\}.
\end{equation}
The constant $c$ is a number of order 1 and shall be kept unspecified.
The electric charge density of fermions is now simply proportional to the
spatial derivative of $f$ and we find the total charge inside a radial
coordinate $z$ to be
\begin{equation}
q(z)=\int dz\, (4\pi\lambda R^2\ J^{0}_{Q})=\frac{1}{\sqrt{2\pi}}
(f-\frac{\theta}{\sqrt{2\pi}})|_{z},
\end{equation}
where we fixed the integration constant by inspecting the electric
energy term in the effective action above. Similarly, the fermion number
inside  $z$ is, up to an additive constant, given by $\eta\sqrt{2/\pi}$
evaluated at $z$.

\vskip 0.5cm
Coupling to  gravity requires further considerations.
First, the effective action above is not manifestly covariant.
The actual two-dimensional metric $g^{(2)}\equiv
\lambda^2\,(dt^2-dz^2)$ has a conformal factor
$\lambda^2$ and the only way to reconcile this with the present form
of the action is to choose the normal ordering scales to be proportional
to $\lambda$. On the other hand, the only two physical mass {\em parameters}
of the effective theory are $e^{\phi}\lambda/R$ and $m_\psi \lambda$,
both of which are proportional to $\lambda$. Hence it is appropriate
to replace $\mu$ by $\lambda\overline{\mu}$, where the specific
choice of $\overline{\mu}$ should not in principle affect the physics.
But in practice, we will study the effective action
at tree level only, which requires a judicious choice of the
ordering scales. For example, we can take
\[ \mu^2= \lambda^2 \overline{\mu}^2 \equiv m_{f} m_{\eta},
\qquad\hbox{$m_{f}$, $m_{\eta}$ are classical {\em masses}
of $f$ and $\eta$, in $(t,z)$ coordinate,} \]
automatically ensuring the covariance of the effective action.
Secondly, there is the matter of zero-point energy, which is irrelevant
before gravity is turned on. To ensure the existence of the cornucopion
for trivial values of the vacuum angle $\theta$, it is neccessary to have
vanishing vacuum energy density, whenever $\cos\theta=1$, both at the
asymptotic infinity and at the other asymptotic region deep inside the
cornucopion.  This can be achieved by adjusting the
fermion mass term to have minimum at zero rather than at $-m_{\psi}\mu
\lambda$. The resulting effective action $S_{\rm eff}$ in an arbitrary
coordinate system is
\begin{equation}
\int dx^2\sqrt{-g^{(2)}}\:\biggl\{\frac{1}{2}(\nabla f)^2
               +\frac{1}{2}(\nabla\eta)^2-\frac{e^{2\phi}}{4\pi R^2}
               (f-\frac{\theta}{\sqrt{2\pi}})^2
                                   -cm_{\psi}\overline{\mu}\,
               (1-\cos\sqrt{2\pi}f\cos\sqrt{2\pi}\eta)\biggr\}. \label{eq:SE}
\end{equation}

\vskip 5mm
\leftline{\bf 3. Vacuum Energy Distribution and the Gravitational Backreaction}
\vskip 5mm

To recover the charge quantization rule, it is sufficient to study the
effective potential  at spatial infinity. In the asymptotic region, the
potential is dominated by the fermionic mass term, whose minima occur at
$f=N\sqrt{\pi/2}$ with $N$ even or odd depending on the asymptotic
value of $\eta$. Therefore,
\begin{equation}
q_{total}=q(z=\infty)=\frac{N}{2}-\frac{\theta}{2\pi},\qquad
\hbox{$N$ is any integer.}
\end{equation}
Obviously the dynamical fermion $\psi$ is responsible for the
new half-integral part, and odd $N$ must correspond to an odd number of
fermions.\footnote{The fermion number $n_{\psi}$ is not conserved
in the presence of a black hole. However, the odd and the even fermion numbers
require different charge quantizations, and $n_{\psi}$ modulo-two is
a good quantum number. It is easy to see that the fermion number $n_{\psi}$
modulo-two is given by $\eta\sqrt{2/\pi}$ evaluated at spatial infinity.}
It is not surprising to find the same results as Callan did \cite{callan},
since the asymptotic form of the effective theory is insensitive to
whether the gravity is turned on or not.

\vskip 5mm
However, the vacuum polarization effect as we approach the black hole can be
very different from the case of a nonsingular monopole in a flat space-time.
Specifically, we want to concentrate on the fermionic ground state
in the background of the cornucopion, a purely magnetic extremal black hole.
Suppose we want to find a parameter region where the noncompact core geometry
of the cornucopion serves as the zero-th order approximation.
Such a configuration would correspond to a narrow and infinite neck of
cornucopion surrounded by harmless and dilute charged cloud of the vacuum
polarization, just as the nonsingular monopole core is, according to Callan
\cite{callan}, surrounded by a harmless and dilute charged cloud of the
vacuum polarization.

A quick look at the effective potential
convinces us that this is not possible unless the fermion mass is actually
zero. Deep inside the neck, the electric energy term is dominant so that
$\sqrt{2\pi} f$ approaches $\theta$, and as a result the ground state
energy density  behaves like
\[ V_{min}\simeq c m_{\psi} \overline{\mu}(1-|\cos\theta|)\: > 0.\]
Since the total vacuum energy inside the throat region is given by integrating
$V_{min}$ along the infinite neck, the gound state built on this background
comes with infinite vacuum energy distributed along the infinite
neck.\footnote{While it is conceivable that some other $\theta$-dependent
effects may cancel $V_{min}$, such a cancellation, if possible, could
occur only for very special values of parameters in the theory. The
$\theta$-independent part is fixed in the previous section by assuming
that the purely magnetic configuration is given by the cornucopion.}

Actually a similar phenomenon occurs for monopoles in a flat space-time,
contributing a vacuum energy which scales like $m_{\psi}^2L_{c}$, where
$L_{c}$ is the distance between the charged shell and the monopole center.
On the other hand, the resulting eletric charge distribution carries
electrostatic energy which scales like $1/R_{c}$, where $4\pi R_{c}^2$ is
the area of the charged shell. In a flat space-time,
$L_{c}\simeq R_{c}$ and the balance between the two contributions fixes the
order of magnitude of $L_{c}\simeq R_{c}$ at $1/m_{\psi}$, as mentioned
earlier.

In a curved black hole geometry, however, $L_{c}$ is now some
measure of the physical distance between the charged shell and the event
horizon, which needs not be proportional to $R_{c}$ (the linear size of
the charged shell) any more. With a cornucopion, in particular, $L_{c}
\rightarrow \infty$ while $R_{c}$ remains finite. It is not possible to
achieve a balance between the two, and it is neccessary to consider
the gravitational backreaction to the vacuum polarization, to understand
the true nature of the fermionic $\theta$-vacua.

\vskip 5mm
To understand the gravitational backreaction,
let us digress a little bit and recall the energetics of the pure cornucopion
solution. The infinite neck of a cornucopion is threaded by a constant flux
of classical magnetic fields. While one would normally expect a uniform
magnetostatic energy density associated with the flux, the energy density
actually vanishes exponentially deep down the neck. The reason is simply that
the electric coupling $e^{\phi}$, inverse square of which appears in the
megnetostatic energy density, is exponentially growing. Note that, in
Einstein-Maxwell theory where the coupling is really a constant, an extremal
horizon forms, hiding whatever divergent behaviour the energy-momentum may
have.

In a sense, the purely magnetic cornucopion of finite ADM mass exists
precisely because the divergent coupling prevents the magnetic energy-momentum
from accumulating divergently. However,
once we turn on the Callan-Rubakov modes with a generic $\theta$,
the energy-momentum given by $V_{min}$ eventually dominates and does
accumulate divergently deep inside the neck. On the other hand, since the
energy-momentum far outside is completely determined by the total
magnetic charge and the fractional electric charge, the ADM mass must be
finite regardless of the vacuum polarization.

Now it is clear what must happen. The gravitational backreaction to the
accumulated effect of $V_{min}$ must eventually create a horizon somewhere
down the would-be cornucopion, rendering $L_{c}$, thus the vacuum energy
contribution to the ADM mass, finite. Hence, the infinite neck must
be terminated by a zero-temperature horizon at finite physical distance.
In such a self-consistent background, one should be able to find the true
vacuum state of the fermion sea.

In fact, one can explicitly check this for small $m_{\psi}$.
In this limit, the geometry near the throat region remains unchanged
since the energy-momentum there is dominated by the magnetic flux, and the
long-neck structure survives until the point where $V_{min}$ is comparable to
the magnetostatic energy density. Then as we travel down the would-be
cornucopion, the dynamics effectively reduce to that of a
2-D dilatonic gravity. The relevant 2-D action can be easily obtained by
dimensionally reducing the complete action (\ref{eq:stotal}) to appear in the
next section.
\[\int dx^2\sqrt{-g^{(2)}}\:\biggl\{e^{-2\phi}\,(-R^{(2)}-4\,(\nabla\phi)^2
+\frac{1}{2\kappa^2})-2V_{min}\biggr\}\]
For reasonable choices of $\overline{\mu}$,
the static solutions of this action can be easily shown to possess two
horizons generically, and the extremal limit thereof corresponds to
a zero-temperature horizon at finite physical distance terminating the long
neck.

\vskip 0.5cm
\leftline{\bf 4. Self-Consistent Geometries and the Fractional Charge }
\vskip 0.5cm

We concluded above that, for generic $\theta$, the gravitational backreaction
to the vacuum energy distribution is always important and that there exists
an extremal horizon stopping indefinite growth of the would-be cornucopion.

In the discussion above however, the fractional electric charge itself
does not seem to play a role as far as the core
geometry is concerned. After all, not only is the charge cloud too large
to approach even the throat region,
but it is known that any electric charge faces an exponentially
divergent potential barrier as it travels down a cornucopion, owing to the
electromagnetic backreaction \cite{alford}. The electric flux tube attached
to the charge costs more and more energy, proportional to the exponentially
divergent coupling squared $e^{2\phi}$, and this tends to push away any
electric charge.

On the other hand, the gravitational backreaction renders this potential
barrier finite, since the coupling cannot be infinite at the regular
extremal horizon, and at least part of the fractional charge should be expected
to be trapped inside the horizon. This last observation raises a question
whether one can explain the newly-formed extremal horizon entirely in terms
of this trapped electric charge.

There are known {\it clean} dyonic black hole
solutions of the dilaton gravity coupled to Maxwell fields \cite{kallosh},
and their extremal limit comes with an extremal horizon.\footnote{In ref
\cite{kallosh}, electric and magnetic charges belong to different $U(1)$'s,
but as far as the static and spherically symmetric geometry is concerned,
it does not matter. One should be careful to distinguish these solutions
from another known family of solutions with an axionic field \cite{dual}.}
Maybe, the vacuum energy density found above seemed so prominent only because
we were using a wrong background. It is a logical possibility
that the self-consistent geometry near the extremal horizon is
dictated by the charges.

In fact, this is precisely what must happen in the large $m_{\psi}$ limit.
As $m_{\psi}$ increase, the density of the charge cloud as well as $V_{min}$
must increase accordingly. The gravitational backreaction to $V_{min}$ creates
the horizon more and more close to the fractional charge cloud which by now
is itself dense enough to distort the geometry. With the increasingly weak
potential barrier, the increasingly massive lump of the fractional charge
will eventually fall into the black hole and the effect of $V_{min}$ will
disappear behind the horizon. Once this happens, an observer outside would be
completely oblivious of the vacuum energy distribution and attribute the
termination of the would-be cornucopion to the fact that
the extremal horizon hides both electric and magnetic charges.

In this section we would like to investigate this possibilty in both the small
and large fermion mass limits, with the latter serving as a consistency check.
The total action dictating the self-consistent geometry is given by the
following.\footnote{In this paper, $\kappa^{-2}$ is the gravitational
constant, while $c=\hbar =1$.}
\begin{equation}
S=S_{\rm eff}-\frac{1}{16\pi\kappa^2}\int dx^4\sqrt{-g^{(4)}}\:e^{-2\phi}
\,(R+4\,(\nabla\phi)^2+\kappa^2F^2). \label{eq:stotal}
\end{equation}
For the purpose of studying the properties of static and spherically
symmetric solutions at the extremal horizon,
which will turn out to be very informative, we can reduce the
field equations to a set of algebraic ones involving various
physical quantities at the  horizon. The key to this simplification is the
regularity of the horizon.\footnote{
In fact, possible mild singularities at the extremal horizon such as
observed by Trivedi\cite{semi} in semiclassical extremal black holes
do not interfere with this derivation.}

If a function of the radial coordinate only is finite and
differentiable at an extremal horizon with respect to a local geodesic
coordinate, some of its covariant derivatives
vanishes there just because $\lambda$ vanish at the horizon.
If we denote the evaluation at the extremal event horizon by the subscript $h$,
\[(\nabla^2f)_{h}=(\nabla f)^2_{h}=0,\qquad \hbox{the same for $\phi$,
$R$, and $\eta$.}\]
As a result, only terms without any derivative of $f$, $\eta$, $\phi$, $R$
survive the evaluation of the static field equations at the extremal horizon.
For instance, combining the dilaton equation and an angular Einstein
equation, we can easily deduce that $R_{h}^{2}=2\kappa^2$, showing that
the transverse size of the neck remains unchanged. On the other hand,
the equation for $\eta$ tells us $\cos\sqrt{2\pi}\eta_{h}=\pm 1$.
{}From some of the remaining equations,
we find two algebraic equations for $e^{\phi_{h}}$ and $f_{h}$.
\begin{eqnarray}
\frac{e^{2\phi_{h}}}{4\pi \kappa^2}(f_{h}-\frac{\theta}{\sqrt{2\pi}})\pm
\sqrt{2\pi}cm_{\psi}\overline{\mu}_{h}\sin\sqrt{2\pi}f_{h}&=&0
\label{eq:eq1} \\
\frac{e^{2\phi_{h}}}{8\pi \kappa^2}(f_{h}-\frac{\theta}{\sqrt{2\pi}})^2
+cm_{\psi}\overline{\mu}_{h}(1\mp \cos\sqrt{2\pi}f_{h})&=&
\frac{e^{-2\phi_{h}}}{4\kappa^2} \label{eq:eq2}
\end{eqnarray}
Now let us consider two limiting cases as promised, to unravel the role of
the fractional electric charge in the formation of the extremal horizon.

\vskip 5mm
When $\kappa m_{\psi} >> 1$, the first equation (\ref{eq:eq1}) tells us
that the value $(\sin \sqrt{2\pi}f_{h})$ is very small and that, for the
lowest energy configuration, the left-hand-side of (\ref{eq:eq2}) is
dominated by the electric energy term $\sim (f_{h}-\theta/\sqrt{2\pi})^2$.
Then, we find the following relation between the coupling and the trapped
charge $q_{h}$  except for integral $\theta/\pi$, when the argument
above breaks down.
\begin{equation}
e^{-4\phi_{h}}=\frac{1}{2\pi}(f_{h}-\frac{\theta}{\sqrt{2\pi}})^2+
O(\frac{1}{c\kappa^2 m_{\psi}^2})=q_{h}^2+ \cdots \label{eq:large}
\end{equation}
This is a nontrivial and significant piece of information, in that this
is exactly what one would expect to be true if the horizon geometry is
completely determined by the electromagnetic charges trapped inside.
It is a matter of straightforward algebra to show that the {\it clean} dyonic
black holes of \cite{kallosh} satisfies $(q/p)^2=e^{-4\phi_{h}}$ with electric
and magnetic charges given by $q$ and $p$. Furthermore, small
$(\sin\sqrt{2\pi}f_{h})$ implies that most of the fractional charge is inside
the black hole, confirming the assertions earlier in this section.

\vskip 5mm

Finally, coming to the small fermion mass limit $\kappa m_{\psi}<< 1$,
we can easily see that the first equation (\ref{eq:eq1}) now predicts very
small $q_{h}\sim(f_{h}-\theta/\sqrt{2\pi})$. Because of this, the
left-hand-side of (\ref{eq:eq2}) is now dominated by the fermion mass
term $\sim m_{\psi}\overline{\mu}$, and this in turn implies the following
characteristic of the self-consistent geometry in the small fermion mass limit.
\begin{equation} |q_{h}|\sim e^{-4\phi_{h}},\quad\hbox{rather than}\quad
   q_{h}^2 \simeq e^{-4\phi_{h}}. \label{eq:small}   \end{equation}
The implication is clear when compared to (\ref{eq:large}).
Now the leading energy-momentum contribution closing off
the infinite neck of the would-be cornucopion (making $e^{-2\phi}$
nonzero) is generated by $V_{min}$ rather than by the trapped electric charge.
Very small amounts of electric charge $q_{h}\sim e^{-4\phi_{h}}$ (not $\sim
e^{-2\phi_{h}}$) are trapped by the newly-formed extremal horizon only
as a secondary effect. The presence of the vaccum energy distribution $V_{min}$
is very real unlike the previous case of large fermion limit.

\vskip 5mm
A couple of remarks are in order. The key formulae (\ref{eq:large}) and
(\ref{eq:small}) are derived without detailed knowledge of $\overline{\mu}$.
All we needed was rough characteristics of it in each limit, such as
$ \overline{\mu}_{h}/m_{\psi} <<e^{2\phi_{h}}$ for small $m_{\psi}$
and $\overline{\mu} \sim m_{\psi}$ for large $m_{\psi}$. This is an important
point because physics should not depend on the choice of the normal
ordering scale, and indeed we managed to isolate such a $\mu$-independent
characterization of the self-consistent geometry in the form of these key
formulae. Another fact we want to mention is  the equations (\ref{eq:eq1})
and (\ref{eq:eq2}) above  show the expected behaviour as $m_{\psi}\rightarrow
0$. Though the precise behaviour does depend on $\overline{\mu}$, the value
$e^{-2\phi_{h}}$  can be
shown to vanish rapidly as $\kappa m_{\psi}$ approaches zero, corresponding to
a longer and longer neck. Eventually when $m_{\psi}$ is identically
zero, the limiting self-consistent geometry is that of a
cornucopion, as it should be.

\vskip 0.5cm
\leftline{\bf 5. Discussion}
\vskip 0.5cm
To summarize, we found that fermionic $\theta$ vacua tends to terminate
the infinite neck of the cornucopion. This fact, by itself, should not be
surprising, since  a nontrivial $\theta$ implies the existence of the
(fractional) electric charge which, if swallowed by the black hole, produces
a {\it clean} dyonic black hole with an extremal horizon. In fact, this is
exactly what happens when $\theta/\pi$ is non-integral and the fermion
mass is sufficiently large. On the other hand, somewhat unexpected is the
behaviour for a small fermion mass. In this case, the dilute charged clouds
hovering far away from the throat region are shown to exist at the cost of
a vacuum energy distribution inside. The energy-momentum associated with
this energy density induces a strong gravitational backreaction, and the
result is again the formation of an extremal horizon at finite distance.
Unlike the case of large fermion mass, however, we found that the charge
penetration to the black hole is at most of secondary effect.
\vskip 5mm

One advantage of using the bosonic form of the matter is, among others,
the exchange of the roles played by the coupling and the mass.
The effective matter action (\ref{eq:SE}) is such that the quantum fluctuation
of the bosons $f$ and $\eta$ are increasingly costly deep down the would-be
cornucopion, and our tree-level estimates of the energy-momentum are
reliable in spite of, or we should say, because of the strong coupling
which is inevitable for small $m_{\psi}$. Even though such a strong coupling
may induce large gravitational fluctuations near the extremal horizon in
small $m_{\psi}$ limit, this should not disrupt our qualitative results.
At most we expect a quantitative modification of the estimate
(\ref{eq:small}).

\vskip 5mm
While we worked with the case of a $SU(2)$ doublet fermion on a unit-charged
would-be cornucopion, similar results should hold for some different models.
The vacuum energy distribution trailing the fractionally charged cloud is
a generic property of the screening and should exist regardless of the model.
What makes this vacuum energy distribution dangerous enough to destabilize
the core geometry is the noncompact nature of the unperturbed core structure,
as emphasized in section 3. Therefore one should expect a similar
destabilization to occur whenever he quantizes massive charged fermion field,
in the background of $\theta$ and an infinite-neck geometry threaded by
constant magnetic flux.

For instance, if we consider a single fermion coupled to $U(1)$ rather than
$SU(2)$, the resulting bosonized matter action must be similar to our own but
involve the charged sector only (corresponding to the $f$ field above, with
$\eta$ frozon out). Since it is the effective potential associated with $f$
which induces the screening and the vacuum energy distribution thereof,
an analogue of $V_{min}$ will appear in the background of a unit-charged
cornucopion. Again the resulting gravitational backreaction will close off
the would-be cornucopion, just as we observed above.

Another interesting case to consider is that of cornucopia of larger
transverse sizes. Since the transverse size is proportional to the magnetic
flux threading the infinite neck, these are highly charged magnetic black
holes. In such a background, one finds analogues of the Callan-Rubakov modes
in the form of zero-modes on the transverse two-spheres. Since the number
of these zero-modes is proportional to the total magnetic flux threading
the two-sphere, we need to deal with not just a pair of two-spinors
$\chi_{\pm}$ but complicated multiplets of the generalized angular momentum
\cite{mono}. Nevertheless, the separation of variables and the dimensional
reduction must be possible, and upon a bosonization trick we expect to find
only two kinds of bosonic fields: $\tilde{f}$, an analogue of $f$ which
keeps track of the electric charge and the vacuum energy
distributions, and $\tilde{\eta}_{a}$'s, the rest of them. There are again
two effective potentials: the electric energy term which is minimized for
vanishing local electric fields and the fermionic mass term which is not.
The upshot is again that the would-be cornucopion develops an extremal
horizon at finite physical distance.

\vskip 5mm
\centerline{\bf Acknowledgement}
\vskip 0.5cm
I am extremely grateful to J. Preskill for his encouragement
and also for sharing his knowledge and insight on the massive
Schwinger model and the $\theta$ vacua. Also I thank S. Trivedi for
illuminating conversations.

\end{document}